\documentstyle[aps, prabib, tighten]{revtex}

\begin{document}


%
\draft

\title
{\bf Remark on ``Conductance and Shot Noise for Particles with Exclusion
Statistics'' by Isakov, Martin, and Ouvry}

\author{
M. P. Das${}^1$ and F. Green${}^2$ \\
${}^1$  {\it Department of Theoretical Physics,
IAS, The Australian National University \\
Canberra, ACT 0200, Australia \\}
${}^2$ {\it CSIRO Telecommunications and Industrial Physics,
PO Box 76, Epping, NSW 1710, Australia}
}

\maketitle

\bigskip
\begin{abstract}
Isakov, Martin, and Ouvry\onlinecite{[2]} have recently proposed a
fresh approach to the potential observation of fractional
exclusion statistics. According to their argument, a clear
signature of fractional statistics should exist in the
shot noise of a Luttinger fluid, an ideal system postulated by some
to underlie the well-founded Laughlin quasiparticle states
of the two-dimensional electron gas.
We elaborate on some delicate points made by Isakov {\it et al.}
and reflect upon the relationship between this novel
intuitive scheme and certain old issues of first principles.


\end{abstract}
\bigskip
\bigskip

The interplay of high magnetic fields and many-body interactions
in a two-dimensional electron gas produces the fractional
quantum-Hall effect (FQHE) at low temperatures. At filling factor
$\nu=1/m$ ($m$ is an odd integer) the Hall resistivity reaches a
plateau, showing that the correlated quasiparticles
carry fractional charge $\nu e$. The fractionally
charged quasiparticles (FCQPs) {\it may} obey fractional exclusion
statistics (FES) \cite{[1]}.

Recent shot-noise and Coulomb-blockade experiments confirm that
one-dimensional FCQPs carry the
edge-state currents in the FQHE,
but cannot establish FES unambiguously.
Isakov {\it et al}. \cite{[2]} consider a novel
theoretical possibility. They generalize the Landauer-B\"uttiker
first-quantized method for one-dimensional
conductance and shot noise to noninteracting particles with
FES, thought to be relevant to the edge-state FQHE.
The explicit formula obtained for the crossover of shot noise to
thermal noise stretches its normal interpretation
in terms of the noninteracting Schottky and Johnson-Nyquist formulas.

The results of Ref. \onlinecite{[2]} may be of experimental interest.
Despite this, we note some potentially troublesome
issues of principle. These warrant closer examination.

(1) To obtain a FES noise spectral density strictly by first
quantization, one must first have a counting method going beyond
the normal occupancies for fermions and bosons. The proposal of
Isakov {\it et al}. fails to recover standard results for normal
fluctuations, as required by second quantization \cite{[3]}.
To redeem their argument they must add an {\it ad hoc} term to
their Eq. (3), maintaining presumed conformity with the
fluctuation-dissipation theorem (FDT).

We draw attention to Isakov {\it et al.}'s stated need to call on
the FDT with no idea whatever of how its microscopic proof
plays out in this new physical context:
a knowledge gap with serious implications.
In every many-body system, the FDT is an absolutely essential
structural link between equilibrium fluctuations and dissipative response.
Its validity must be deduced from the model's axioms;
it can never be established inductively.
This means that while the microscopic fluctuations prescribe
the observable conductance, the converse does not hold.
The present approach is unable to derive the
appropriate FDT. It allows only speculation on some
FDT-like ansatz that is cut and fit, inductively, to one's
notions of what the fluctuations might be (given the conductance).
Such {\it ad hoc} inversion of the theorem's prescriptive
meaning severely undermines the microscopic credibility of the
state-counting argument.

(2) FCQPs carry the FQHE edge-state current in {\it one dimension},
and thus represent the excitations of a correlated Luttinger fluid.
This is totally unlike a fluid of ``independent'' quasiparticles.
There are no grounds, in second quantization, to describe exclusion
statistics by an ansatz that interpolates intuitively
between the limiting occupancies for {\it free} bosons and fermions.
Not surprisingly, the oversimplified accounting leads again to a
familiar difficulty: a term must be added {\it ad hoc}
to keep faith with the FDT. For bosons a further
unphysical feature appears, in that the
zero-frequency spectral density for shot noise diverges.
This dilemma is avoidable -- in  second quantization -- via
the generalized commutation relations introduced long ago
by H. S. Green \cite{[4]}.

(3) The crossover of the spectral density from thermal to shot
noise is derived by assuming $\exp(\beta \mu) \gg 1$. This
occurs when either the thermal energy (temperature) $\beta^{-1}$
is very small, or the chemical potential $\mu$ is very large.
The latter corresponds to a high-density system. If the FCQPs are
truly independent, the system will be ballistic.
In the quantum ballistic limit there is no shot noise.

Shot noise is inherently a nonequilibrium phenomenon; any
{\it correlations} in it must involve nonlinearity in the applied field.
For normal fermions, linear theories based on quantum-coherent
(or on semiclassical)
diffusion imply a smooth crossover from low temperatures and
high fields, to high temperatures and low fields.
While this has strong appeal in making sense of experiments,
a first-principles nonequilibrium theory is not in sight.

The Landauer-B\"uttiker approach is akin to Kubo linear
response. For shot noise in correlated systems,
a major question should be how its suppression
is modified by many-body interactions. In a strictly
single-particle, linear-transport picture,
suppression enters via the factor $T(1-T)$ where $T$ is the
one-particle transmission probability through the system.
In both limits $T \to 0$ and $T \to 1$, shot noise tends to zero.
For $T \to 0$ the system is nonconducting;
trivially, there is no shot noise. When $T \to 1$, the system
is ballistic; again, within strictly first-quantized
treatments of fluctuations, there is no shot noise.
However, as the quasiparticles of a conventional Fermi
system {\it do} interact, it is fair to wonder whether
their shot noise is indeed so comprehensively suppressed.

One is now asked to go beyond the normal, and to postulate
fractional statistics for (nearly free) quasiparticles
in the Luttinger liquid, an utterly correlation-dominated
and certainly unconventional many-body system.
Isakov, Martin, and Ouvry\cite{[2]} raise many more
open problems than have been answered satisfactorily
for noise in the Landau model of normal quasiparticles.
Furthermore, their concept of correlated
quasiparticle systems amounts to the following.

\begin{itemize}
\item
Strongly correlated systems can always be renormalized into
an ensemble of quasiparticles. These may possess fractional
statistics (which can be intuited).

\item
Since the low-lying excitation spectrum of the system will
resemble one-body behavior, one can suppose that the 
residual quasiparticle interactions are literally weak.
The system is effectively free.

\item
One can then safely analyze the two-body fluctuations
(noise) of such systems as if the quasiparticles
did not actually interact.
Said differently: the mean-square fluctuations of
a correlated system are fully and uniquely given by the effective
mean single-particle properties.
\end{itemize}

In our view, such a proposal hardly fits what is known even
for normal Fermi liquids. One need only recall that the low-order
Landau parameters (essentially associated with quasiparticle
renormalization) are quantitatively locked into all the
higher-order ones (associated with the interactions
{\it among} the quasiparticles) by a standard sum rule
\cite{[5]}.
This basic constraint will draw in higher and higher
angular moments of the ``residual'' quasiparticle
interaction as the coupling
strength in the underlying Hamiltonian is increased. No-one is
yet sure, of course, what holds for the Luttinger case.



\end{document}